# Wide-band frequency modulation of a terahertz intrinsic Josephson junction emitter of a cuprate superconductor

M. Miyamoto[1], R. Kobayashi[1], G. Kuwano[2], M. Tsujimoto[3], and I. Kakeya[1,*]

[1]Department of Electronic Science and Engineering, Kyoto University, Nishikyo, Kyoto 615-8510, Japan

[2]Advanced Manufacturing Research Institute, National Institute of Advanced Industrial Science and Technology (AIST), Tsukuba, Ibaraki 305-8564, Japan

[3]Global Research and Development Center for Business by Quantum-AI Technology (G-QuAT), National Institute of Advanced Industrial Science and Technology (AIST), Tsukuba, Ibaraki 305-8568, Japan

[*] Corresponding author: kakeya@kuee.kyoto-u.ac.jp

## Abstract

**Communication using terahertz (~$10^{12}$ Hz) electromagnetic waves is critical for developing 6th-generation wireless network infrastructures. Conflictions between stable radiation and the modulation frequency of terahertz sources impede the superposing of transmitting signals on carrier waves. The Josephson junctions included in a cuprate superconductor radiate terahertz waves with frequencies proportional to the bias voltages. Thus, the modulation of the bias voltage leads to the modulation of the Josephson plasma emission (JPE) frequency. This study aims to demonstrate the generation of frequency-modulated (FM) terahertz continuous waves from Josephson junctions. We achieved that 3 GHz sinusoidal waves were superimposed on 840–890 GHz carrier waves, and the modulation bandwidth up to 40 GHz when a JPE was utilised. The results verify that the instantaneous JPE frequency follows the gigahertz-modulated bias voltage. The wide-band FM terahertz generation by a monolithic device shows a sharp contrast to the mode-lock frequency comb constructed by highly sophisticated optics on a bench. A further increase of the modulation amplitude facilitates up- or down-frequency conversion over more than one octave. The obtained FM bandwidth exhibited an improvement of two orders of magnitude in the demodulation signal-to-noise ratio compared to the amplitude-modulated signal. The demonstrated FM-JPE stimulates further research on terahertz communication technology and metrology using superconducting devices.**

## Introduction

The generation and detection of electromagnetic waves in the terahertz range have been extensively investigated over the past 20 years[1–3]. In particular, coherent radiation and detection in the sub-terahertz region, which are crucial for communications, have recently attracted attention as a key technology for next-generation ultrahigh-speed wireless communication networks[4,5]. To implement a terahertz source for a communication infrastructure, sophisticated modulation technologies to superimpose the information to be propagated are required. However, research in this area has yielded little progress because there is often a trade-off between stable monochromaticity and the coherence of radiated terahertz waves in the modulation process. In the case of resonant tunnelling diodes (RTD), which are currently the most advanced source for demonstrating terahertz communication, mono-bit digital communication using amplitude shift keying has been employed[6]. Research on analogue or digital communications using angular modulation, including frequency and phase[7], has rarely been conducted. Compared to amplitude-modulated (AM) communication, frequency-modulated (FM) communication has significant advantages in terms of a high demodulation signal-to-noise ratio (SNR).

A Josephson junction comprises a pair of superconducting electrodes with a normal barrier layer in-between, serving as a converter between a DC voltage and an AC superconducting current across the barrier[8], and is employed as a voltage standard utilizing the Shapiro step. In a high-Tc cuprate superconductor, forming a stack of intrinsic Josephson junctions (IJJs)[9,10] leads to the synchronized oscillations of Josephson currents among numerous IJJs. This synchronisation results in the outward radiation of electromagnetic waves at terahertz frequencies[11], which we refer to as Josephson plasma emission (JPE)[3,12,13] because the standing transverse Josephson plasma waves inside a cuprate superconductor induce the radiation via surface oscillating current[14–16]. The frequency of the radiated electromagnetic wave is proportional to the applied voltage per IJJ, according to the Josephson relation[17]. JPE has been observed over a wide frequency range[18] corresponding to changes in the voltage for a single device[17], with detection outside the cryostat from 0.5 to 2.4 THz[19] or inside the same cryostat from 1 to 11 THz[20]. Frequency tuning from 0.15 to 1 THz[21], covering whole future 6G band, makes liquid-helium free JPE operations[22] more beneficial. The outlook of a space-time crystal provides a unique insight for condensed-matter physics[23].

If this frequency variation follows a temporal voltage variation, modulation of the bias voltage of the JPE results in an FM terahertz source. Increasing the modulation amplitude expands the frequency bandwidth, thus creating an ultrawideband FM terahertz source as a monolithic device. However, it has been pointed that the frequency variation is partially attributed to superconducting material parameter changes via effective temperature variation in asymmetric devices[18,24]; thus, the generation of FM terahertz waves when fast (> GHz) bias modulation is applied to JPEs is not apparent. Meanwhile, FM radiations of low-$T_c$ Josephson junctions have been demonstrated. In a Nb/NbO$_x$/Pb single Josephson junction, FM signals centred around 9 GHz have been obtained under megahertz modulated biases [25]. Recently, a comb-like spectrum with a bandwidth of 4 kHz centred at 5.6 GHz and with a laser action has been reported in an Al/AlO$_x$/Al superconducting quantum interference device (SQUID) [26]. These findings have developed into superconducting receivers[27] utilised for radio telescopes[28], Josephson parametric ampliphier[29], and so on.

This report presents the results obtained from analysing electromagnetic radiation induced by biasing sinusoidal signals with frequencies of 10 kHz and 2 – 3.5 GHz superimposed on approximately 2 DC volts, using a high-resolution Fourier transform spectrometer. The analysis tells that the JPE radiation frequency follows the temporal evolution of bias voltage through the ac Josephson effect with negligible time delay. When the modulation frequency exceeds the spectrometer resolution, an FM spectrum described by a comb-like Bessel-function sum is verified, where the comb-tooth spacing corresponds to the modulation frequency. The temporal evolution of a thousand stacked IJJs was analysed to establish in-phase synchronization by comparing the centre frequency evolution of the spectra with the modulation frequency. Furthermore, the radiation spectrum at 3 GHz modulation is observed beyond the frequency range of the unmodulated-bias radiation, which is different from the frequency comb in laser operation[30] given by the product of the medium amplification factor and the resonance mode. This implies that wide-band (> 10 GHz) FM terahertz transmitters can be developed from a high-temperature superconductor single crystal using uncomplicated fabrication and microwave techniques. JPE devices possess unique capabilities, such as the monolithic generation of circularly polarised continuous waves[31,32] , narrow linewidths of less than 50 MHz [33] despite having a broad tuning range[34], synchronous radiation from multiple mesas achieving radiation power up to 0.6 mW[35,36], and high radiation-power efficiency [37]. This unique FM terahertz emitter illuminates a new route towards the 6G wireless network with existing THz devices. Our findings can facilitate the realization of sub-THz frequency combs with bandwidths more than one octave and FM-THz radar[38] for next-generation metrologies.

## Results

### Microwave (~ GHz) modulation and their amplitude evolution

Figure 1a shows a typical radiation spectrum for a mesa voltage with a superimposed sinusoidal signal of power $P_m = 25$ dBm and a frequency of $f_m = 3$ GHz. The spectrum was obtained by doing fast Fourier transform to an interferogram shown in Figure 1b, in which beat features are found. The bias-modulation schemes are illustrated in Figure 1c and the measurement setup is depicted in Figure 1d,e[39]. In the spectrum, 15 peaks were identified at frequencies centred at $f_c = 856$ GHz and equally separated by $f_m = 3$ GHz; thus, peak frequencies are $f_c + nf_m$ with $n = 0, \pm 1, \cdots \pm 7$. Their linewidths were approximately identical to and limited by the frequency resolution of the spectrometer of 0.8 GHz. Figure 1a includes the ideal FM spectrum represented by

$$I_{\mathrm{FM}} = \frac{1}{2} A_c^2 \sum_{n=-\infty}^{\infty} J_n^2(m_f), \tag{1}$$

as a series of bars, where $J_n(x)$ is an $n$-th order Bessel function of the first kind and the modulation index is $m_f = 5.7$. $A_c$, originally defined as the amplitude of the carrier waves at a frequency $f_c$, was chosen to reproduce the experimental peak intensity ratio. See Supplementary Note 1. The excellent agreement to the spectrum clearly indicates the demonstration of a terahertz FM-JPE. Taking the SNR at the $n = +6$ peak ($B = 36$ GHz) as $S/N = 12$, the data transfer rate is $C = B \log_2(1 + S/N) = 130$ Gbit/s based on the Shannon-Hartley theorem of digital communication, where $B = 2(m_f + 1)f_m$ is the bandwidth. The demodulation SNR is improved to $3(B/2f_m)^2 \cong 100$ times with respect to AM transmission, exhibiting a two-order magnitude increase in the effective radiation power in comparison to AM-based terahertz communications.

The spectrum obtained by varying $P_m$ from 25 dBm to $-5$ dBm is shown in Figure 2ab. As $P_m$ is decreased, the bandwidth $B$ of the observed spectrum decreases, and at $-5$ dBm, the spectrum becomes a familiar unimodal spectrum at approximately 860 GHz. Given that $B$ is proportional to $V_m$ through $m_f$, the measurement circuit impedance $Z$ including the device and cable at 3 GHz can be estimated from the slope obtained for the $\sqrt{P_m}$ dependence of $B$ (Figure 2c). From the slope of the linear fitting, the impedance of the transmission line including the contacts is estimated to be as high as $Z \sim 3$ M$\Omega$, considering the mesa impedance of 510 $\Omega$ derived based on the current-voltage characteristics shown in Figure 2d. This large $Z$ may be attributed to electrical connections by silver paste on the sample substrate. In addition, Figure 2e shows the spectra when $f_m$ varies between 2 and 3.5 GHz. The peak spacing is equal to $f_m$. This electrically controlled continuous change in $f_m$ is a unique characteristic of the FM-JPE, in contrast to comb generations using a quantum cascade laser[40,41] and RTD[42], wherein the device geometry and the external resonator determine their comb separations. In this case, the modulation bandwidth decreases with increasing frequency. This is because the modulation amplitude applied to the mesa decreases for the same synthesiser output power owing to the increased losses and reflections in the bias cable. The details are provided in the Supplementary Note 2.

The integrated spectrum intensity $S(\omega_c) = \int_{\omega_L}^{\omega_U} I(\omega) \, d\omega$ decreases considerably with an increase in the modulation amplitude, as shown in Figure 2c (right axis), where $I(\omega)$ is the (experimental) intensity spectrum, and the upper and lower limits of the integration, $\omega_U$ and $\omega_L$, are sufficiently far from the spectrum. This is consistent with the radiation intensity measurement (Figure 2f), wherein the detected intensity decreased for the modulated bias. We argue that this decrease is mainly owing to the decrease in the unmodulated spectrum intensity $S_{\mathrm{um}}(\omega_c)$, accompanied by the deviation of the instantaneous frequency from the maximum radiation frequency. Another possibility is the polarisation change owing to bias modulation when multiple resonance modes exist in close proximity in the radiation region. Given that the spectrometer is equipped with a polariser at the entrance of the interferometer shown in Figure 1d, the intensity of the introduced terahertz wave is reduced if the bias modulation affects not only the radiated frequency but also the radiated polarisation [43]. However, this is unlikely in this case, because an identically designed device[44] radiates highly linear polarisation at an axial ratio of up to 15 with little bias dependence.

Figure 3a shows spectra with different mesa voltages $V_{\mathrm{mesa}}$while maintaining a bias modulation of $f_m = 3$ GHz and $P_m = 25$ dBm. Compared to the spectra without modulation, shown in Figure 3b, the spectra are symmetrically broadened in the frequency domain by the 3-GHz modulation. The peak spacing is also maintained at $f_m$, resulting in a spectral shift parallel with respect to the mesa voltage, corresponding to the centre frequency $f_c = \omega_c/2\pi$ of the spectrum. The height of the modulated spectrum is maximum with $f_c$ close to the maximum frequency of $S_{\mathrm{um}}(\omega)$. This indicates that the spectrum intensity is determined by $f_c$ and is less relevant to the reduction of the radiation intensity at the ends of the unmodulated radiation frequency range, exhibiting a sharp contrast to the case of $f_m = 10$ kHz shown in Figure 3c, where and the spectral shape strongly depends on $f_c$. The intensity ratio of the peaks is shown as a coloured symbol in Figure 3d, where the $+|n|$th peaks tends to be higher than the $-|n|$th peaks at a lower $f_c$ and vice versa. This reflects the intensity distribution of the unmodulated bias radiation.

Temperature evolution

By decreasing the device temperature to 18 K, the radiation intensity with DC bias increased up to twice of that at 25 K while voltage for the intensity maximum slightly increases, as depicted in Figure 4ab. The unmodulated radiation frequency versus the mesa voltage overlapped with the data at 25 K. The monotonous increase in radiation intensity at lower temperatures contrasts with the previously reported non-monotonous temperature dependence of the spectral intensity in an identically designed device [44]. It is plausible to consider that a slight difference in the device structure, such as the connection between the mesa and the triangular patches, improves the robustness of radiation polarisation at low temperatures. Thus, the expected decrease in quasiparticle dissipation loss becomes apparent in the present work. Similar FM spectra were obtained between 18 K and 27 K using gigahertz bias modulation, as data at 18 K shown in Figure 4c.

A further increase in temperature resulted in the appearance of one-order weaker radiation between 40 and 65 K. At 50 K, the unmodulated radiation frequency is 465 GHz, as shown in Figure 4d. This is a unique radiation attributed to the pair of triangular patches, as previously reported [44], because no corresponding Josephson plasma standing wave exists along the mesa short edge of 50 μm and radiation obtained from a standing wave along the long edge is mostly reflected by the linear polariser of the interferometer. More intriguingly, the modulation of 3 GHz-25 dBm for the 465-GHz mode did not yield a comb-like spectrum, instead exhibited a broadened spectrum, as depicted in Figure 4d. This inconsistency is presumably due to a violation of coherence among the stacked IJJs, influenced by the bath temperature. A minor peak was found at the lower-frequency side of the unmodulated spectrum and did not change with the 10-kHz modulation. This implies that the excitation of multimode and incoherent standing waves inside the device, including the patches, is consistent with the linewidth measurements, wherein a significantly broadened spectrum with a linewidth of more than 1 GHz was observed at 40 K at the low bias region [33].

## Discussion

To highlight the unique bias evolution of the 3-GHz FM spectra, a comparison with the spectra under slower-bias modulation is discussed. For $f_m = 10$ kHz shown in Figure 3c, radiation was observed in the same frequency range as in the unmodulated case. However, there is an essential difference compared to the 3-GHz modulation case in addition to the unresolvable comb separation. In the case of low-frequency modulation, for mesa voltages $V_{\mathrm{mesa}} \approx 2.0$ V, near the lower limit of the radiation region, the frequencies associated with locally prominent spectral intensities are visible. In some cases, it is significantly asymmetric and qualitatively different from the 3-GHz modulation case. This is because, in the gigahertz-modulation case, the coherent oscillating current at $f_c$ (= 840–890 GHz), induced by the AC Josephson effect, is modulated, thus exhibits a spectrum close to an ideal FM. However, when modulated at 10 kHz, the instantaneous voltage determines the instantaneous Josephson frequency $\dot{\theta}(t)$ (Eq. (S1) in Supplementary Note 1), resulting in the superposition of the unmodulated spectral intensities by a time-domain distribution. To investigate this phenomenon in more detail, the spectra with sinusoidal and triangular modulations at 10 kHz are shown in Figure 5a. The frequency expansion, i.e., the separation between the two main peaks for the triangular modulation, is reduced to 83 % compared to the sinusoidal modulation. The Fourier series of a triangular wave with amplitude $V_m$ and period $\frac{2\pi}{\omega_m}$ is $\frac{8V_m}{\pi^2} \sum_{k=1}^{\infty} \frac{(-1)^{k-1}}{(2k-1)^2} \sin(2k-1)\omega_m t$.

The coefficient of $\sin \omega_m t$ corresponds to $0.81V_m$; thus, the higher-order components of $k \geq 2$ do not contribute to the bandwidth. Since the amplitude of the $3\omega_m$ component is $3^{-2}$ of the amplitude of the $\omega_m$ component, another pair of peaks separated by $18.9 \times 3^{-2} = 2.1$ GHz exists. However, the expected peak separation is too small to be resolved by the spectrometer, although tiny peaks around the centre imply the existence of the $3\omega_m$ component. This is indicative of not only a pure sinusoidal signal, but also a weighted sum of sinusoidal signals, which can represent arbitrary analogue information, and can be transmitted using FM-JPE. Additional modulation of 3 GHz and 25 dBm (thicker curves) completely smears the pair of peaks in the 10-kHz modulation, indicating that no coherent FM is established.

These experimental results show that for the 10-kHz modulation, no radiation is observed when the modulated instantaneous frequency $\dot{\theta}(t)$ exceeds the radiative frequency range for the unmodulated case regardless of the centre frequency $f_c$. However, for the 3-GHz modulation, radiation may be observed when $f_c$ is in the radiation range for the unmodulated bias, regardless of the modulation width. This is compared at the low-frequency side, where the decrease in $S_{\mathrm{um}}(\omega_c)$ is more significant. In Figure 5b, no spectrum intensity was observed below 836 GHz for the 10-kHz modulation, whereas a peak at 820 GHz was observed for the 3-GHz modulation. Moreover, $f_c$ of the lowest observed spectrum with the 3-GHz modulation was slightly above 836 GHz (Figure 5b upper). Thus, when $f_c$ corresponds to the radiation frequency range with an unmodulated bias, a spectrum is detectable regardless of the modulation bandwidth. If higher-frequency modulation is applied with a sufficient amplitude $V_m$, it is expected to produce a spectrum with a frequency range well-above the unmodulated radiation range.

The maximum possible FM bandwidth was estimated by the experimental results. Taking the intensity distribution of $S_{um}(\omega_c)$ into consideration, the FM spectrum intensity $S_{FM}(m_f, \omega_c)$ may decrease the power of $m_f$ for large $m_f$ as shown in Supplementary Figure 5b. Therefore, the experimental $S_{FM}(m_f, f_c = 856\ \mathrm{GHz})$ obtained from the data in Figure 2c can be extrapolated to give $m_f \sim 100$. As a result, we found that the spectrum of the half-octave modulation $m_f = 71$, i.e., $B = 432\ \mathrm{GHz}$, could be resolved with an SNR>3. More details are described in the Supplementary Note 3. JPE devices with broader radiation range[34] are capable to demonstrate wider FM bandwidth.

Next, the synchronisation between stacked IJJs is discussed. Assuming that a relaxation time $\tau_{sync}$ is required to develop or decay the synchronous oscillations of the stacked IJJs, the maximum amplitude of the surface macroscopic current (corresponding to $A_c$ in Eq. (1)), induced by synchronisation, is delayed by $\tau_{sync}$ during the application of the radiation mesa voltage. This is schematically illustrated in Figure 6. However, the frequency of the surface current followed the mesa voltage $V(t)$ with a negligible delay. This implies that the instantaneous voltages of the stacked IJJs were kept equal for the gigahertz modulation. For $f_m \gg \tau_{sync}^{-1}$, it is possible to maintain the synchronous oscillation while the instantaneous radiation frequency $\dot{\theta}(t)$ exceeds the original synchronising region. $\tau_{sync}^{-1}$ is relevant to the phase correlation between adjacent IJJs[45] induced by the capacitive coupling[46,47] and the phase relaxation time of the IJJs that characterise the retrapping rate of the switching dynamics from superconducting to resistive states[48,49]. According to Kuramoto's phase dynamics theory of non-linear oscillators, the coupling between IJJs demonstrates a decay in synchronisation for weakly coupled oscillators with slightly different natural frequencies. This decay is characterised by a time constant that is anticorrelated with the sum of their frequency difference and a $2\pi$ cyclic coupling function between the two oscillators[50]. Considering two neighbouring IJJs, the frequency difference arises from the *ab*-plane geometry difference because of the trapezoidal cross section of the mesa device, and the coupling function is attributed to the charging of the superconducting layers owing to Josephson tunnelling with a radiation frequency. This two-IJJ problem[45] can be developed into a thousand-IJJ problem as the origin of the inertia of synchronisation, which requires an extensive modification of the retrapping phenomena observed in a single IJJ. To estimate the $\tau_{sync}$ value of an IJJ stack, the analysis of the continuous $f_m$ evolution of the comb-like spectrum at one end of the radiation frequency region is necessary because the asymmetry of the spectrum is partly attributed to the frequency distribution of the radiation intensity for the unmodulated bias.

Figures

Figure 1: Comb-like spectrum with 3 GHz and 25 dBm modulation and experimental setup.
**a,** Radiation spectrum (line-connected symbol) at T = 25 K, mesa voltage $V_{\mathrm{mesa}}$ = 2.05 V, superimposed modulation frequency 3 GHz, amplitude 25 dBm. The ideal FM spectrum for a carrier frequency of 856 GHz, modulation frequency of 3 GHz, and modulation index of 5.7 is shown with blue bars. **b,** Interferogram for the spectrum. **The upper panel** corresponds to a mirror scan length of approximately 16 cm. **The lower panels are** partially enlarged views approximately 0.25 $f_m^{-1}$ apart in time delay. **c,** Schematic of bias (frequency) modulation (rainbow-coloured sinusoidal curve) and current-voltage characteristics (spheres at bottom plane; same data as Figure 2d) coloured by radiation intensity. **d,** Arrangements of the JPE device, optical parts, and the bolometer, which are identical to Ref [39].The crystal structure of Bi2212 shown in the bottom exhibits radiation direction of terahertz waves with respect to crystallographic axes. **e,** Device connection for modulated biases and definition of device voltage $V_{\mathrm{dev}}$, including mesa voltage $V_{\mathrm{mesa}}$ and non-linear contact voltages mentioned in Supplementary Note 4. **f,** Measured device. Four triangular antenna patches and a long rectangular electrode with a short tab made of silver are attached to the Bi2212 mesa with proper epoxy and SiO₂ insulations.

Figure 2: Amplitude and frequency dependence of spectra with microwave bias modulations and basic device properties.
**a,** Spectral transitions for the variation of the modulation amplitude at a $V_{\mathrm{mesa}}$ = 2.05 V and a superimposed modulation frequency of 3 GHz, decreasing from 25 dBm to -5 dBm at the synthesiser output. **b,** Extracted spectra at $P$ = 25, 15, 0 dBm (indicated by broken holizontal lines in a) from top to bottom. **c,** Modulation amplitude dependence of spectrum bandwidth (left axis) and integrated intensity (right axis). The bandwidth is determined by 80% of the integrated intensity between 829 and 871 GHz. Integrated intensity at the maximum modulation amplitude decreases to half that of the unmodulated case. **d,** Current-voltage characteristics. Blue plots represent the unmodulated bias, red plots indicate 3 GHz, 25 dBm modulation. The current-voltage characteristics do not change with the modulation. **e,** Comparison of radiation intensities with and without voltage modulation. The integrated intensity of the spectrum is also slightly reduced because of the increase in the modulation amplitude. **f,** Spectra at various modulation frequencies. The modulation frequency was varied, and the synthesiser output was constant at 25 dBm. Upper axis and vertical grids show frequency difference from the spectrum center.

Figure 3: Centre frequency (mesa voltage) evolution of the spectra.

**a,** Radiation spectra for $f_m$ = 3 GHz, $P$ = 25 dBm modulation at various mesa voltages indicated in the legend. No peak height inhomogeneity or asymmetry is observed in each spectrum, and the overall intensity of the spectrum seems to be determined by the radiation intensity at the centre frequency $f_c$. **b,** Unmodulated spectra at various mesa voltages at 25 K. The frequency resolution is three times larger than that of modulated bias cases. Two faint peaks positioned 20 GHz away from the main peak are attributed to the window functions. **Inset of b,** Mesa voltage dependence of radiation frequency. From the slope, the number of IJJs contributing to the radiation is estimated as 1050. **c,** Mesa voltage dependence of the radiation spectrum at 10 kHz, 0.1 V modulation. The symmetry of the spectrum depends on $f_c$, unlike that for the 3 GHz modulation. **d,** The eight highest peak frequencies with respect to the centre of the spectra as a function of $f_c$, which is determined by the average of the six highest peak frequencies $f_{1st}, \cdots f_{6th}$. Symbol size scales to peak height, and symbol colour corresponds to the intensity ratio relative to the highest peak of the spectrum. The trend of the peak height from the highest to lowest value is $\Delta f = \pm 4 f_m, \pm f_m, \pm 5 f_m, \pm 3 f_m$. The intensity ratio is larger for positive $\Delta f$ when $f_c$ is lower, which is similar to the trend for the unmodulated bias radiation intensity.

Figure 4: Temperature dependence of radiation properties.

**a,** Maximum radiation power as a function of device temperature (red broken). Bolometer responses with sweeping both mesa voltage (colour) and temperature (horizontal) are plotted. **b,** Current-voltage characteristics (upper) and radiation detection (lower) at 18 (blue) and 25 (green) K. **c,** 3 GHz modulated spectra with $V_{mesa}$ = 2.05 V obtained at 18 K. Centre frequency and spectrum intensity increased and peak-width is slightly broadened in comparison to data at 25 K. **d,** Spectrum at T= 50 K and $V_{mesa}$ = 1.11 V with and without modulations of 3 GHz and 10 kHz. The unmodulated spectrum is not unimodal and broadened by the 10 kHz modulation. A comb-like structure is not readily observed in the 3-GHz modulated spectrum.

Figure 5: Comparison of 10 kHz and 3 GHz modulations.

**a,** Comparison of sinusoidal and triangular modulation at 10 kHz, 0.1 V. The separation values between the two main peaks at $V_{mesa}$ = 2.06 V are 22.8 and 18.9 GHz for the sinusoidal and triangular modulations, respectively. The thick line represents the spectrum for the case where the 3 GHz, 25 dBm modulation is also superimposed, and no comb-like peaks are observed. **b,** Spectral low-frequency regions at 3 GHz (upper) and 10 kHz (lower) modulation at mesa voltages shown in the legend. Downward arrows point to the lower limits of the radiation for the two modulation frequencies.

Figure 6: Schematic description of synchronization among the stacked IJJs for low-frequency modulation and high-frequency modulation.
Assuming that the radiation intensity (broken curve) is delayed for $\tau_{\text{sync}}$, for the duration wherein $V(t) \propto \dot{\theta}(t)$ (solid curve) is outside the radiation range (orange shadow) shorter than $\tau_{sync}$, there is no reduction in the temporal average of the radiation frequency bandwidth (thick solid arrow) as in **a**. For $f_m \ll \tau_{\text{sync}}^{-1}$ shown in **b**, FM bandwidth significantly changes whether $V(t)$ exceeds the radiation region or not (grey). **I-III,** Schematic illustrations of fully synchronized (**I**), partially synchronized (**II**), and incoherent (**III**) Josephson plasma waves (red-wavy planes) excited in the stacked IJJs. Thickness of an orange conical arrow represents the amplitude of the macroscopic surface current oscillating as a result of the synchronization. Labels **I-III** in **a** and **b** indicate corresponding synchronization status during the modulation periods.

## Acknowledgement

I.K. thanks S. Dhillon, J. Tignon, Y. Uzawa and Y. Irimajiri for discussions on terahertz spectroscopy. G. K. and M. T. express their gratitude to Y. Kaneko for technical assistance with sample fabrication. This work was supported by the Japan Society for the Promotion of Science (JSPS) KAKENHI [grant nos. JP20H02606 (to I.K.), JP19H02540 (to M.T.)] and ISHIZUE 2023 of Kyoto University (to I.K.).

## Author contributions

I.K. and M.T. conceived and designed the experiments. M.M., R.K., and I.K. performed the measurements, analysed and interpreted the data. M.T. and G.K. designed and fabricated the JPE device. I.K. and M.M. wrote the paper with input from R.K. and M.T. All authors participated in reviewing and revising the manuscript.

## Competing interests

The authors declare no competing interests.

## Methods

### Device preparation and radiation detection

The mesa structure and its wiring silver electrodes were fabricated using photolithography and argon-ion milling of a $Bi_2Sr_2CaCu_2O_{8+\delta}$ (Bi2212) single crystal[12] grown using the traveling floating zone method[51], as shown in Figure 1f. The mesa structure has a width of 50 μm, length of 300 μm, and thickness of 1.5 μm, which are slightly different from those of the device in Ref.[44]. The wiring electrodes were connected to an external circuit using electrode pads formed on a $7 \times 7$ mm$^2$ sapphire substrate. Unmodulated and modulated biases were applied from outside of the cryostat. The current was measured using series-connected resistors, while the mesa voltage and the power of the radiated electromagnetic waves were measured via voltage terminals and a silicon bolometer (Infrared Lab.), respectively. The intensity of the radiated terahertz waves was chopped at a frequency of approximately 200 Hz, close to the window facing to the JPE device, and the bolometer output was amplified using a preamplifier (SR560) and a lock-in amplifier (SR850). Current, voltage, radiation, and delay (for spectroscopy) data were acquired using a PXI data acquisition system and analysed using LabVIEW VIs.

Figure 2d shows the current-voltage characteristics of the device, and Figure 2f shows the mesa voltage $V_{\text{mesa}}$ dependence of the bolometer-detected electromagnetic wave intensity for unmodulated and 3-GHz modulated biases. The intensity peak for the modulated case is smaller. This is due to averaging radiation intensity as a function of radiation frequency. The electromagnetic radiation is observed at $V_{\text{mesa}}$ between 1.98 and 2.18 V, with a maximum intensity at approximately 2.05 V. In this voltage range, all the IJJs that form the mesa structure are considered to contribute to the radiation in the voltage state. As shown in the inset of Figure 3b, the radiation frequency measured for an unmodulated bias is proportional to the mesa voltage. Based on its slope, the number of IJJs contributing to the voltage was determined to be 1050, which is slightly larger than the IJJ number of 1000 estimated from the 1.5 μm height of the mesa structure as measured using an atomic force microscope. This is presumably owing to the presence of 50 IJJs under the silver electrodes, which are called contact mesas. The planar mesa geometry facilitates the estimation of the radiation frequency by assuming a refractive index $n_r = 4.2$ [52]. The half-wavelength mode along the short edge ($w = 50$ μm) corresponds to $f_c = \frac{c}{2n_r w} = 714$ GHz, which is considerably lower than the main-spectrum frequency. Possible origin of this discrepancy is discussed in Ref. [44].

## High-frequency superimposed biasing circuit

Silver wires with a diameter of 50 $\mu m$ and a length of a few millimetres were electrically connected to the electrode pad using silver paste and the other ends of the wires were soldered to the terminal on the cold finger of the cryostat. A semi-rigid coaxial cable was connected to the terminal using LakeShore's low-temperature stainless-steel coaxial cable and an SMP connector. The other end of the semirigid coaxial cable was connected to the outside of the cryostat using a hermetic SMA connector, to which a high-frequency bias voltage was applied. A bias tee (Mini-Circuit ZX85-12G-S+) was used to superimpose the microwave modulation bias voltage. A synthesiser (HP83624B) was connected to the RF input side of the bias tee for 2– 4 GHz microwaves and a function generator (WF1967) was connected to the DC input side for DC to 100 MHz signals. A schematic of this connection is shown in Figure 1e.

High-resolution Fourier transformation spectrometer

In this study, a laboratory-built Martin-Puplett Fourier transform interferometer[39] with a frequency resolution of 0.8 GHz for a 20 cm scan of the movable mirror was used to observe the comb-like spectrum. The longest scan takes approximately 20 mins. For the Fourier transform analysis of the acquired interferogram, an FFT filter and appropriate window (e.g., Hanning) functions were used to eliminate experimental artefacts.

Although the spectrometer used in this study is adequate for the comb-like peak separation caused by the modulation process, it is necessary to further increase the frequency resolution to accurately identify the modulation index and evaluate the single-peak line widths. For this purpose, mixing experiments using a heterodyne mixer with a local oscillator[53], which have been previously used in JPE research[33,54,55], are necessary.

## Data Availability

All of the data that support the findings of this study are reported in the main text, supplementary information. Source data are available from the corresponding authors on reasonable request.

## Methods-only references

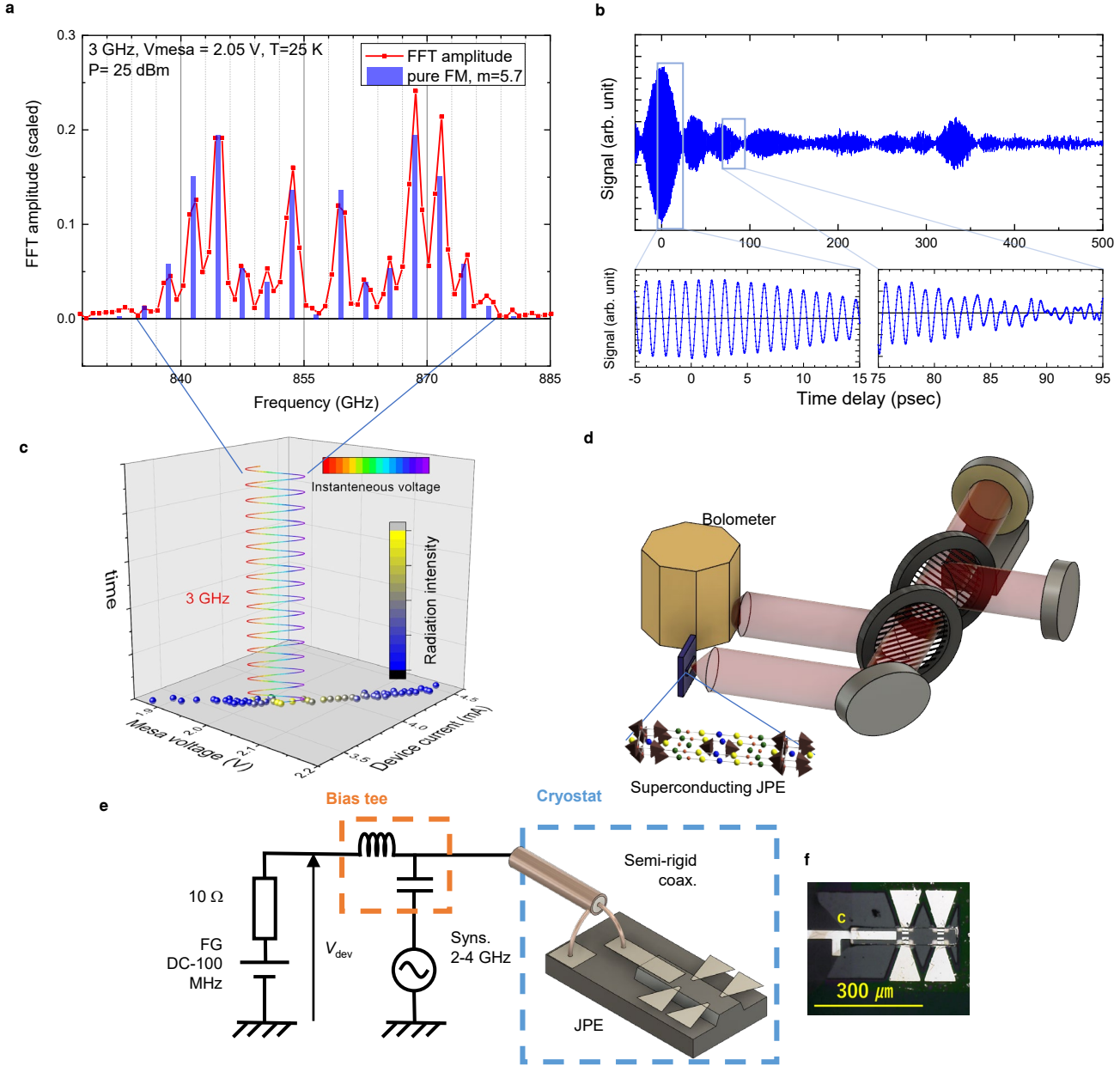

Figure 1

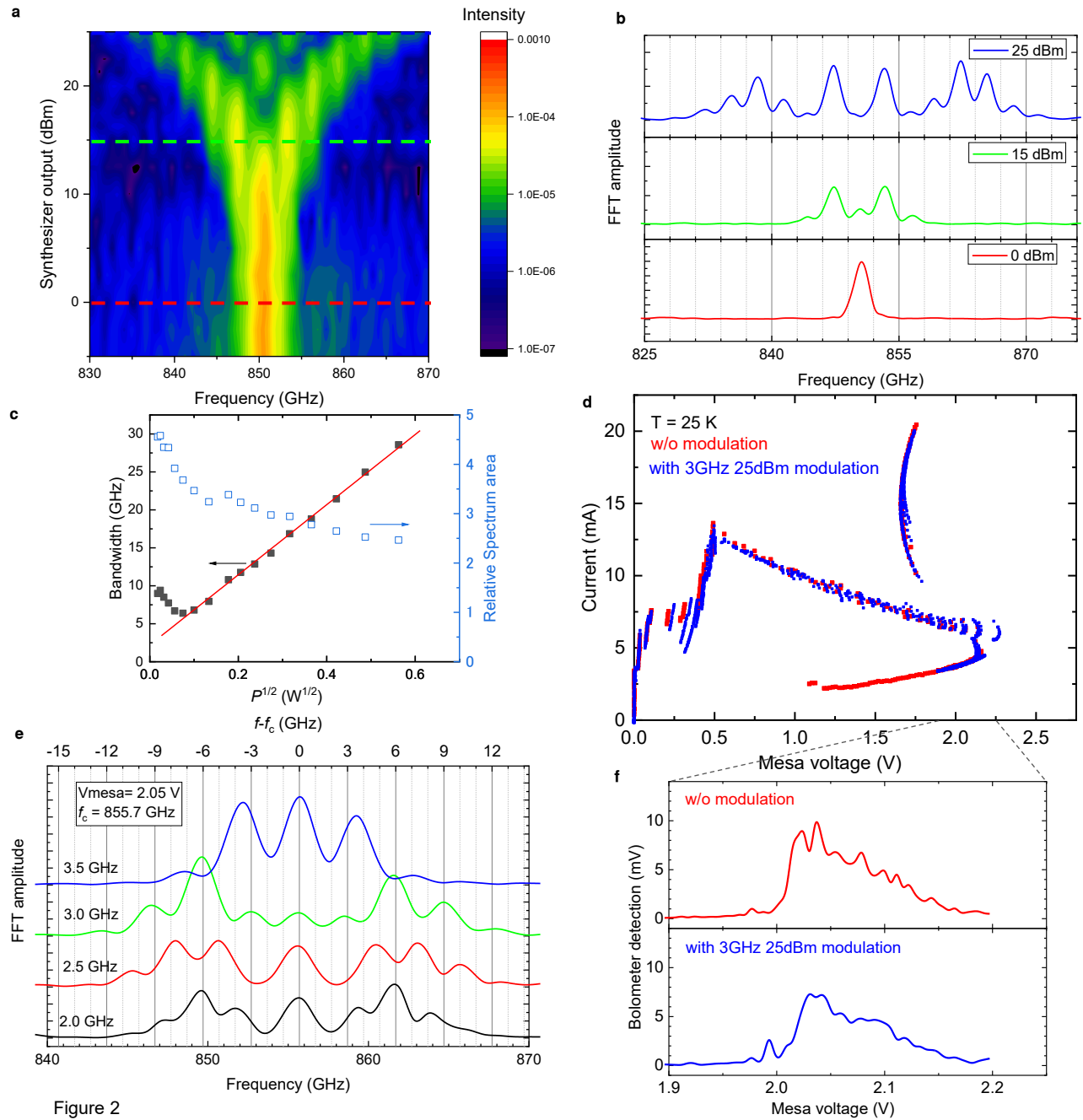

Figure 2

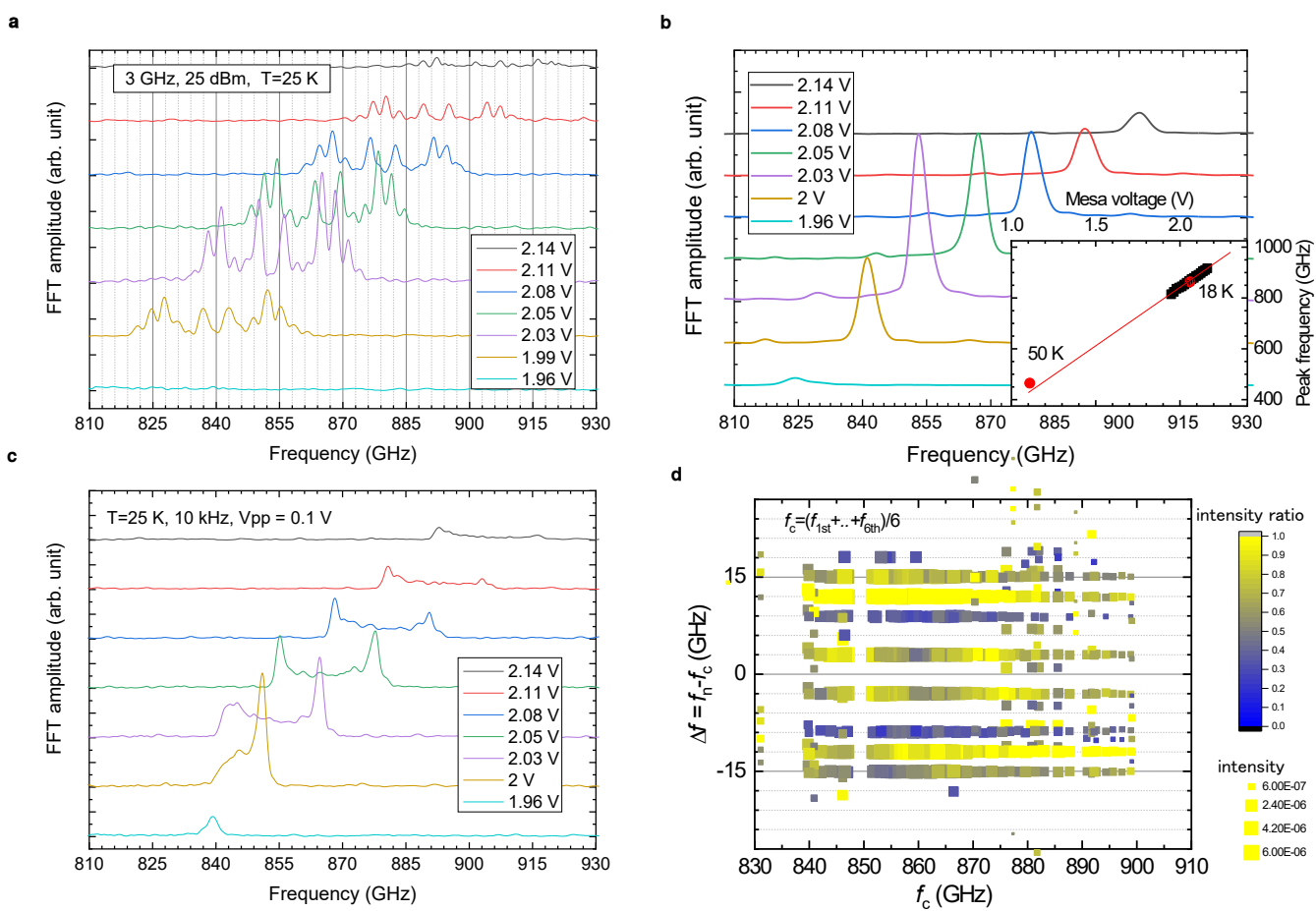

Figure 3

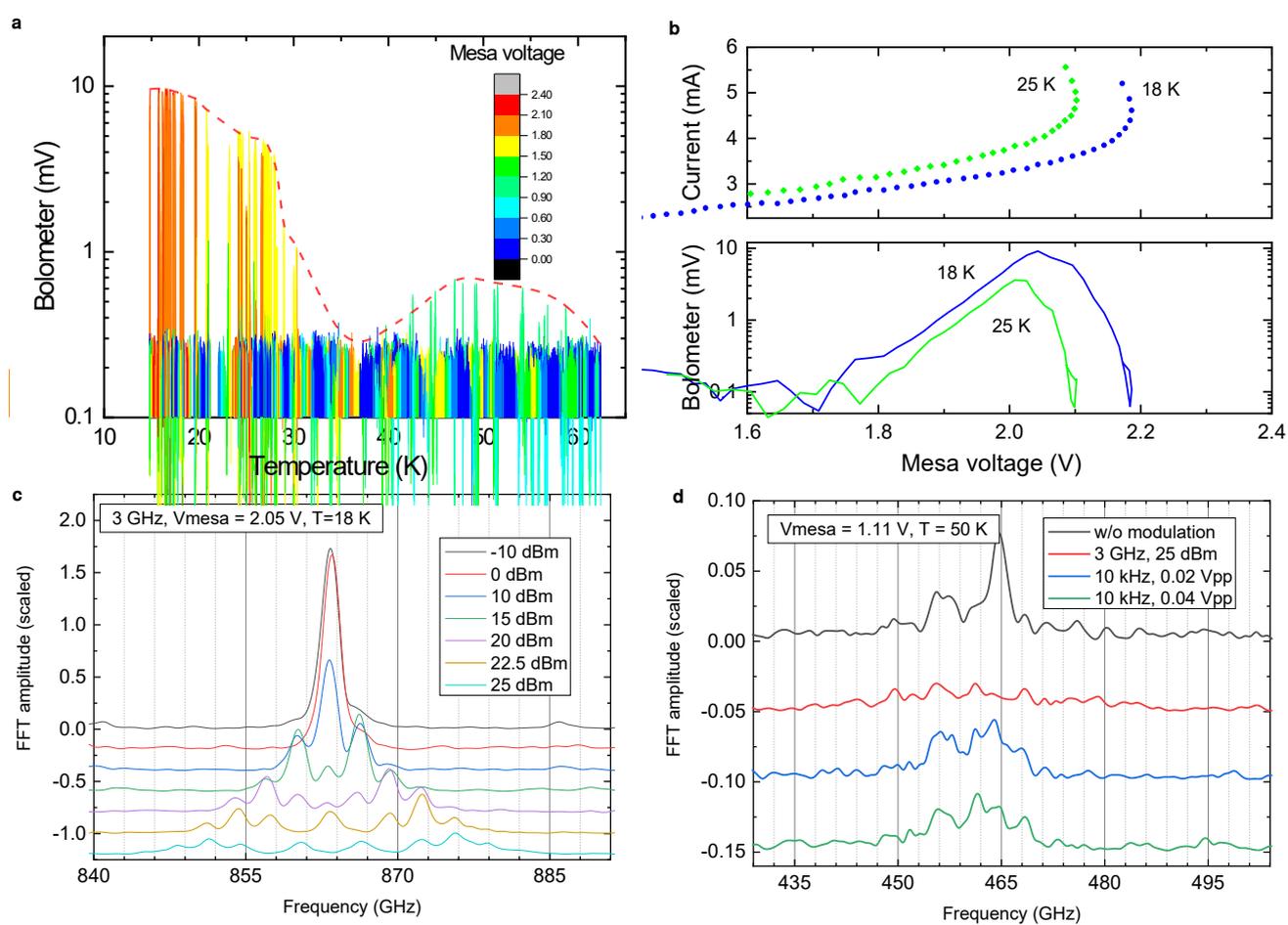

Figure 4

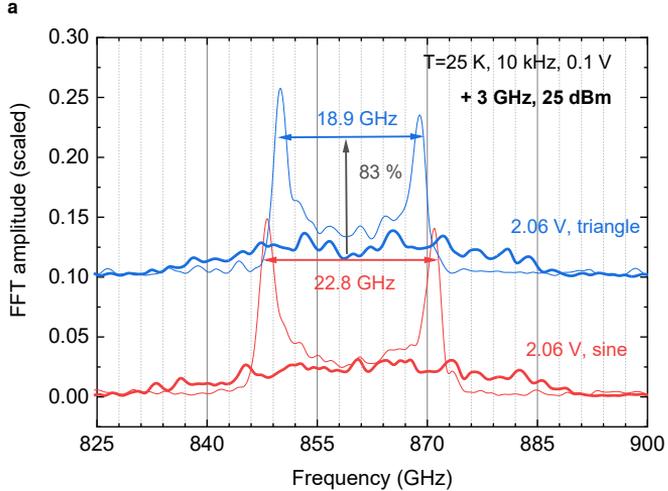

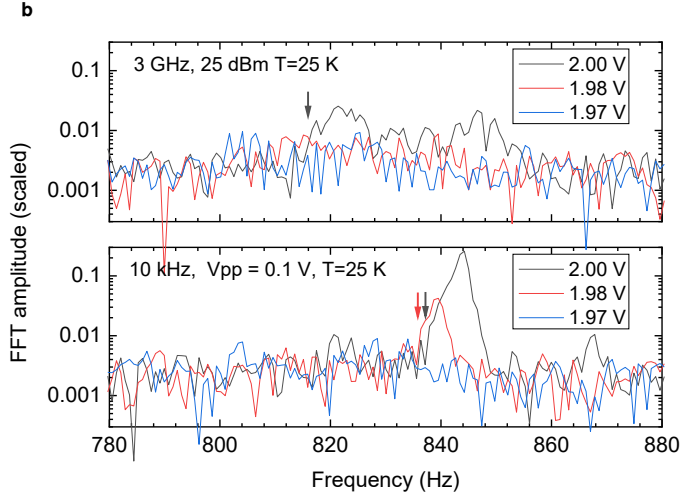

Figure 5

**a: 3 GHz**

**b: 10 kHz**

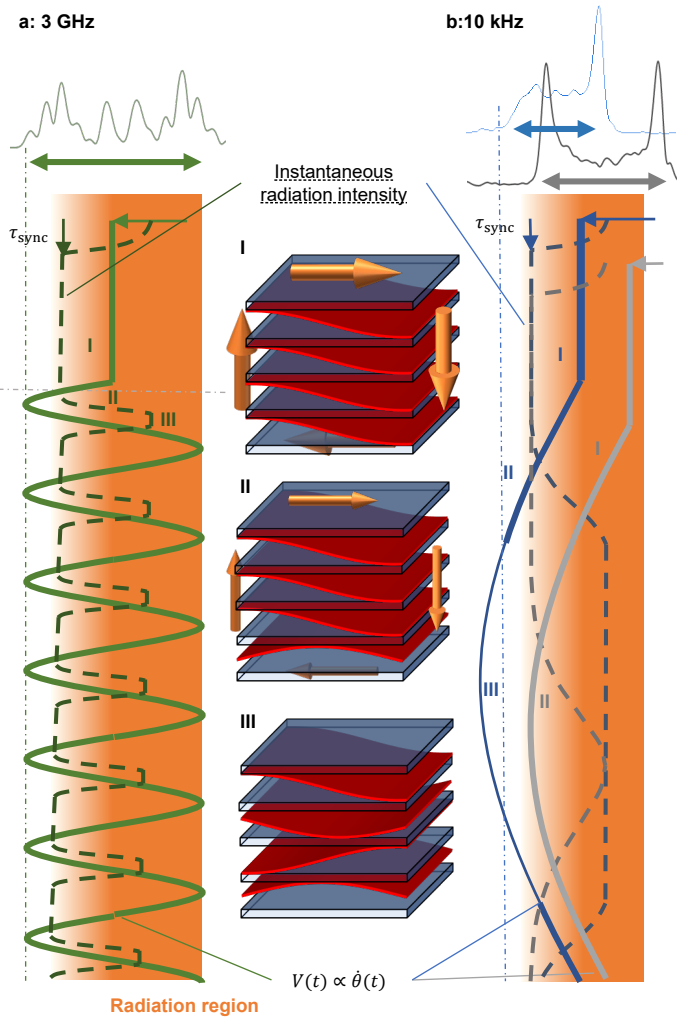

Instantaneous
radiation intensity

$\tau_{sync}$

I

II

III

$\tau_{sync}$

I

II

III

$V(t) \propto \dot{\theta}(t)$

**Radiation region**

Figure 6

# Contents of Supplementary Information

## Supplementary Note 1. FM Josephson oscillation of an IJJ stack

In a single Josephson junction with the gauge-invariant phase difference $\theta$ between the superconducting electrodes, $I = I_c \sin\theta$ and $V = \frac{\hbar}{2e}\frac{d\theta}{dt}$ hold for tunneling current $I$ and biased voltage $V$, where $I_c$ is the maximum tunneling Josephson current, $\hbar$ is the Dirac constant, and $e$ is the elementary charge. Now, let us consider the case applying a time-dependent voltage $V(t) = V_0 + V_m\cos\omega_m t$ to a structure composed of N-stacked IJJs. $V_0$ is the DC voltage, $V_m$ is the amplitude of the modulating voltage, $\omega_m$ is the angular frequency of the modulation voltage. In this case, the instantaneous gauge-invariant phase difference $\theta(t)$ is the integral value of the applied voltage as,

$$\theta(t) = \frac{2e}{\hbar}\int\frac{V(t)}{N}dt = \frac{2eV_0}{N\hbar}t + \frac{2eV_m}{N\hbar\omega_m}\sin\omega_m t \qquad (S1)$$

Equation (S1) is equivalent to the instantaneous angle of the FM electromagnetic wave with the center angular frequency $\omega_c = \frac{2eV_0}{N\hbar}$, modulation index $m_f = \frac{2eV_m}{N\hbar\omega_m}$, and modulation angular frequency $\omega_m$, and if the AC Josephson current is radiated as it is as an electromagnetic wave, the signal $s(t) = m_f\sin\omega_m t$ can be propagated. The intensity spectrum of the FM electromagnetic wave is expressed by the following,

$$I_{\mathrm{FM}}(\omega_n) = \frac{1}{2}A_c^2\sum_{n=-\infty}^{\infty}J_n^2(m_f), \qquad (S2)$$

where $A_c$ is the amplitude of the carrier wave and $J_n(x)$ is an $n$-th order Bessel function of the first kind with $n$ being an integer. From this, the power spectrum of the FM signal has peaks with heights of $J_n^2(m_f)$ at angular frequencies of $\omega_n = \omega_c + n\omega_m$. As shown in Supplementary Figure 1, the spectrum has multiple peaks in a comb-like pattern at intervals of $\omega_m$ centered at $\omega_c$ and its bandwidth is determined by $m_f$. The envelope of the spectrum has maximum intensities near $\omega_c \pm (m_f - 1)\omega_m$ and decreases slowly on the side toward $\omega_c$ and decreases sharply away from $\omega_c$. In other words, when the spectrum corresponding to Eq. (S2) is obtained, the communication at $f_c = \omega_c/2\pi$ with the bandwidth of

$$B = 2(m_f + 1)\omega_m/2\pi, \qquad (S3)$$

is demonstrated. As practical values, applying $V_0 =$1.6 V with $V_m = 0.037$V and $f_m = \omega_m/2\pi = 3$ GHz to a mesa of 1000 IJJs give rise to generate FM signal with $f_c = 800$ GHz with the bandwidth $B = 42$ GHz, as plotted in **Supplementary Figure 1**.

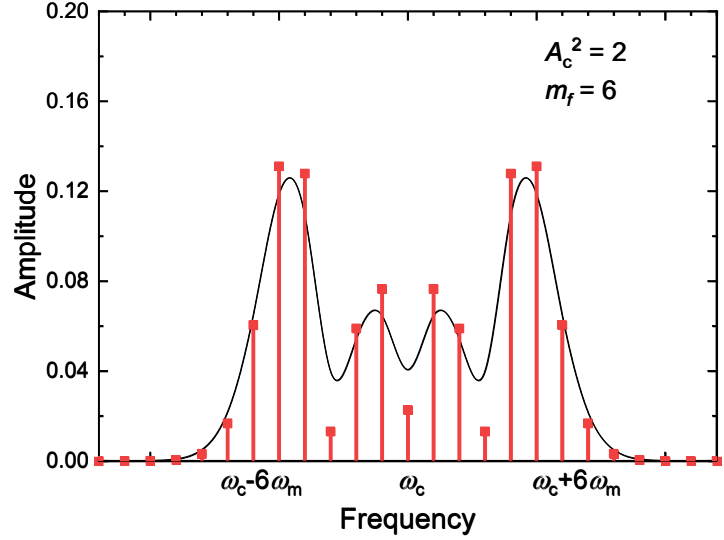

**Supplementary Figure 1: Ideal FM spectrum for $m_f = 6$.** The solid curve is a spline connecting the peak of comb teeth, which simulates data taken by low-resolution spectrometer.

## Supplementary Note 2. Calculation of transmission loss

From the Kirchhoff's current law on the equivalent circuit for the transmission line depicted in Supplementary Figure 2,

$$\tilde{I}_s(t) = \tilde{I}_C(t) + \tilde{I}_m(t), \tag{S4}$$

where $\tilde{I}_s(t) = I_s e^{i\omega_m t}$, $\tilde{I}_C(t) = I_C e^{i\omega_m t}$, and $\tilde{I}_m(t) = \tilde{I}_m e^{i\omega_m t}$ denote complex source, leak, and modulation currents, respectively. The voltage law is

$$\tilde{V}_s(t) = -L\frac{d\tilde{I}_s}{dt} + R\tilde{I}_s(t) + \tilde{V}_m(t), \tag{S5}$$

where $\tilde{V}_s(t) = V_s e^{i\omega_m t}$ and $\tilde{V}_m(t) = \tilde{V}_m e^{i\omega_m t}$ are source and modulation voltages, and $L$ and $R$ are inductance and resistance of the transmission line, respectively. These two equations are reformulated as

$$V_s = (-i\omega_m L + R)(i\omega_m C\tilde{V}_m + \tilde{I}_m) + \tilde{V}_m, \tag{S6}$$

where $C$ is the capacitance of the transmission line. The radiating JPE device, which is a series connection of a thousand of IJJs oscillating synchronously, can be considered as a single RCSJ circuit. Since we here focus on current at frequency of $\frac{\omega_m}{2\pi}$ which applies a finite bias to the JPE, the current flowing across the shunted junction can be zero. Then, the current conservation law in the JPE part is given by

$$\tilde{I}_m = (i\omega_m C_{\mathrm{JPE}} + R_{\mathrm{JPE}}^{-1})\tilde{V}_m, \tag{S7}$$

where $C_{\mathrm{JPE}}$ and $R_{\mathrm{JPE}}$ are capacitance and resistance of the JPE device. From Equations (S6) and

(S7), we obtain

$$V_s = \left\{ \omega_m^2 LC' + \frac{R}{R_{\text{JPE}}} + 1 + i\omega_m \left( RC' - \frac{L}{R_{\text{JPE}}} \right) \right\} \tilde{V}_m, \qquad (S8)$$

where $C' = C + C_{\text{JPE}}$ is the coupled capacitance of the system. Therefore, the reciprocal of the product of modulation index and modulation angular frequency $(m_f \omega_m)^{-1}$ is proportional to

$$\frac{V_s}{V_m} = \left| \frac{V_s}{\tilde{V}_m} \right| = \left( \omega_m^2 LC' + \frac{R}{R_{\text{JPE}}} + 1 \right)^2 + \omega_m^2 \left( RC' - \frac{L}{R_{\text{JPE}}} \right)^2 = c_4 \omega_m^4 + c_2 \omega_m^2 + c_0, \qquad (S9)$$

where $c_4 = (LC')^2$, $c_2 = 2LC' + (RC')^2 + \left( \frac{L}{R_{\text{JPE}}} \right)^2$, and $c_0 = \left( \frac{R}{R_{\text{JPE}}} + 1 \right)^2$. Using the total impedance of the circuit $Z_{\text{total}} = Z + Z_{\text{mesa}}$ mentioned in Main, the equation

$$V_m^{-1} = \frac{c_4 \omega_m^4 + c_2 \omega_m^2 + c_0}{\sqrt{2 Z_{\text{total}} P_m}}, \qquad (S10)$$

holds. **Supplementary Figure 3** shows that experimentally obtained $V_m^{-1} = 2e/(hNB/2)$ does not have a parabolic $f_m^2$ dependence, suggesting that $Z_{\text{total}}$ strongly depends on $f_m$ in this region.

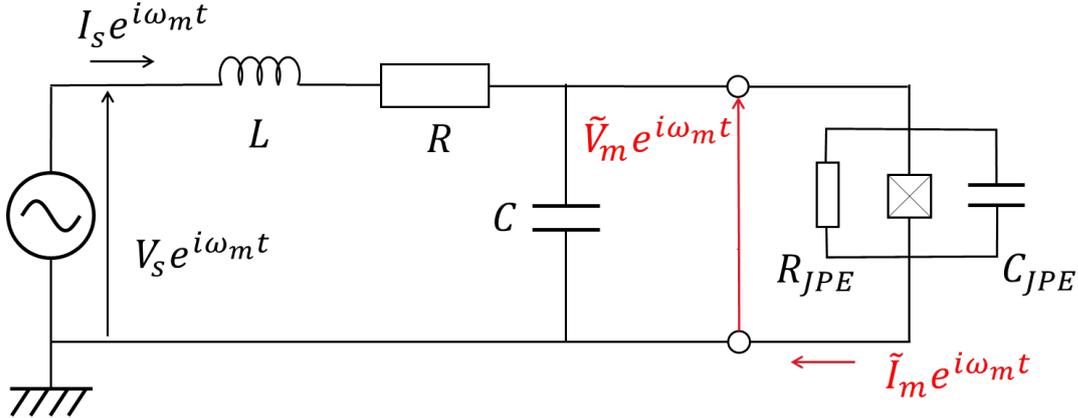

**Supplementary Figure 2: Equivalent circuit of the measurement circuit.** $\tilde{V}_m e^{i\omega_m t}$ is voltage applied to the IJJ stack.

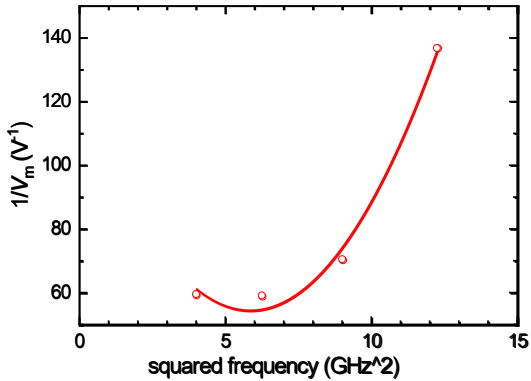

**Supplementary Figure 3:** $V_m^{-1}$ derived from the measured bandwidth *B* versus

squared modulation frequency $f_m^2$.

## Supplementary Note 3. Possible maximum bandwidth

Detectable maximum bandwidths have been estimated in the following manners. First, $A_c$ in Eq. (S2) is determined from the experimental unmodulated spectrum height ($P_m = -5$ dBm in Figure 2a of the main text) with $m_f = 0$. Then an ideal FM spectrum is drawn with increasing $m_f$ while keeping the determined $A_c$. Supplementary Figure 4 is the comparison between the ideal FM spectrum with $m_f = 142$, $f_m = 3$ GHz, and $f_c = 852$ GHz. The highest peaks at 441 and 1264 GHz are several-times higher than the floor noise level, thus one-octave modulation is possible. Note that this argument works under the assumption the reduction in integral intensity with increasing modulation amplitude discussed in the main text is negligible.

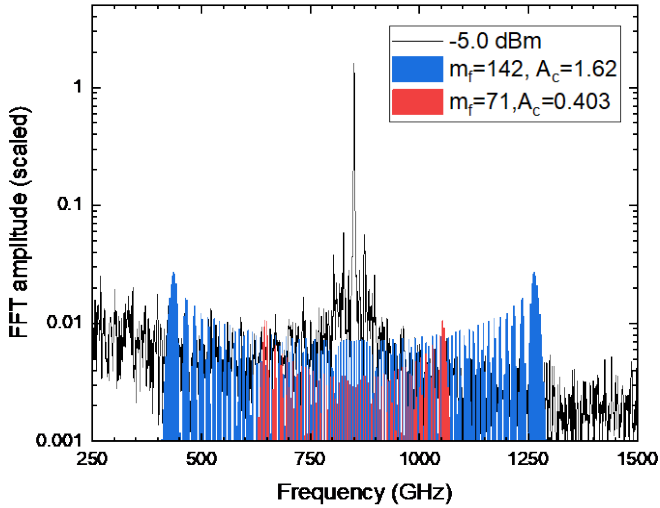

**Supplementary Figure 4: Comparison between experimental spectrum and virtual one(half)-octave FM spectra.** Blue is for $m_f = 142$ and $A_c = 1.62$ with neglecting the intensity decrease derived from Eq. (S2) and red is for $m_f = 71$ and $A_c = 1.62$ with taking the intensity decrease into consideration. Here $f_m = 3$ GHz, and $f_c = 852$ GHz are not changed.

Next, let us estimate FM spectrum intensity $S_{\mathrm{FM}}(m_f, \omega_c) = \int_{\omega_L}^{\omega_U} I(\omega)\, d\omega$ with taking the frequency dependence of radiation intensity into consideration. Assuming that $S_{\mathrm{FM}}(m_f, \omega_c)$ is proportional to the sum of temporal distribution of unmodulated radiation intensity $I_{\mathrm{um}}(\omega)$ per a sinusoidal frequency modulation

$$\omega = \omega_c + m_f \omega_m \sin \omega_m t, \tag{S11}$$

with a period of $T_m = 2\pi/\omega_m$. For a parabolic frequency dependence of $I_{\mathrm{um}}(\omega) = I_0 - a(\omega - \omega_0)^2$ in the vicinity of $\omega_0$, the intensity maximum, we obtain

$$S_{FM}(m_f) \propto \langle I_{\text{um}}(\omega) \rangle_t = T_m^{-1} \int_0^{T_m} I_{um}(t)dt = I_0 - a\left\{\frac{m_f^2 \omega_m^2}{2} + \Delta\omega_0^2\right\}, \qquad (S12)$$

where $\Delta\omega = \omega_c - \omega_0$. The modulated spectrum intensity also shows a parabolic $m_f$ dependence. To estimate $S_{FM}(m_f)$ for a large $m_f$, experimental $S_{FM}(P_m = 25 \text{ dBm}, \omega_c)$ is used instead of $I_{um}(\omega)$ in Eq. (9). Supplementary Figure 5a shows temporal variation of the expected radiation intensity for a bias modulation period. $S_{FM}(m_f, \omega_c)$ is considered to be equivalent to the average of the plot for $m_f$, thus the normalized $S_{FM}(m_f, \omega_c)$ decreases with increasing $m_f$ as shown in Supplementary Figure 5b. Experimental values evaluated from Figure 2c are also plotted. Although the measured reduction of the spectrum intensity is larger within the experimental $m_f$ range, the measured reduction rate at larger $m_f$ is smaller than the simulated result. For the half-octave modulation of $m_f = 71$, the linear extrapolation in the log-log plot implies that spectrum intensity decreases to 0.25 of the spectrum intensity at $P = -5$ dBm. Under this assumption, the highest peak of the half-octave modulated spectrum is almost twice as the noise level because $0.4 A_c$ is used to estimate the modulated spectrum height as shown in Supplementary Figure 4. Therefore, half-octave ($B{\sim}400$ GHz) FM will be the largest limit of wideband transmission in this case.

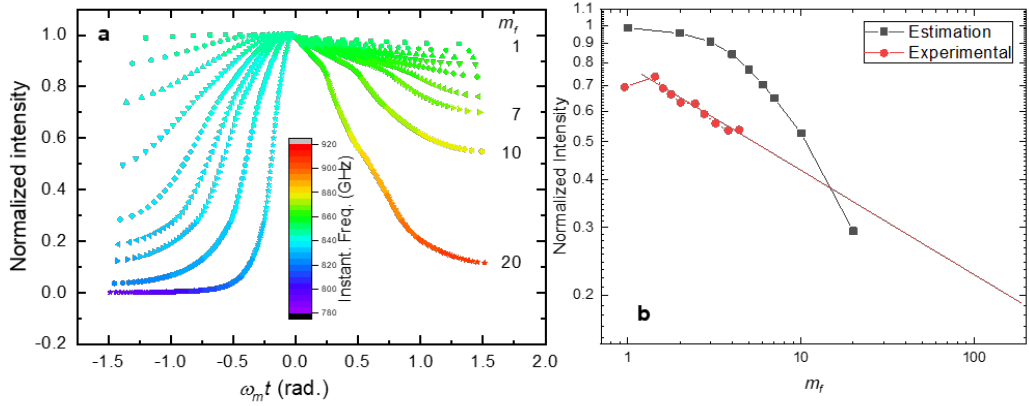

**Supplementary Figure 5: Expectation of expanded spectrum intensity. a**, Intensity distribution per a modulation period $2\pi/\omega_m$ estimated by center frequency dependence of spectrum intensity. With increasing $m_f$, instantaneous frequency $\dot{\theta}(t)$ expands further form the center frequency, then above $m_f = 10$, $\dot{\theta}(t)$ goes away from the radiation range. **b**, Estimated spectrum intensity and experimental spectrum intensity as functions of $m_f$. The experimental $m_f$ is estimated from the bandwidth $B$ plotted in **Figure 2c** with adjusting the $B$-interception of the linear fitting to satisfy $B = 2(m_f + 1)f_m$. The red line is the guide for a linear extrapolation.

## Supplementary Note 4. Empirical reduction of contact resistance

To evaluate voltage applied to the series of IJJs, voltage contributions due to wiring inside and outside the cryostat and contact resistance have to be removed because voltage terminals are connected on the way of bias line outside the cryostat. Linear fittings to the current-voltage characteristics in the vicinity of the origin, where all IJJs are considered to be zero resistance, have been usually employed so far. However, the reduction often turns to be more overestimated at higher current region probably because of a non-ohmic contact resistance between the top IJJ and the evaporated electrode. Here, we suggest to apply an antisymmetric non-linear function of

$$V_{\text{lead}}(I) = c_0 \operatorname{atan}\left\{\frac{p(I - I_0)}{\pi}\right\} + bI + dI^3,\tag{S13}$$

to the lowest voltage (closest to the $V = 0$) branch of the I-V characteristics. **Supplementary Figure 6** shows that the agreement of the lowest branch and the fitted function is excellent. The mesa voltage is the reduction of $V_{\text{lead}}$ from the device voltage $V_{\text{dev}}$ defined by **Figure 6 d**, whose values are collected by data acquisition system.

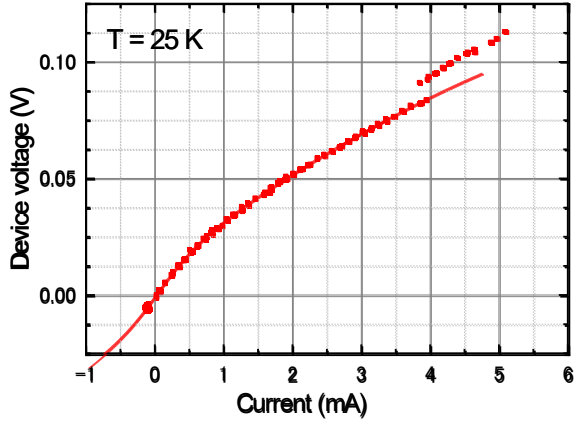

**Supplementary Figure 6: Enlarged current-voltage characteristics of the device and fitting to extract mesa voltage.**